\documentclass[prx,10pt,aps,bibnotes,twocolumn,superscriptaddress]{revtex4-2}

\usepackage[pdftex]{graphicx}
\usepackage{amsbsy,amssymb,amsmath,bm,mathtools}
\usepackage{amsfonts}

\usepackage{xspace}
\usepackage{bm}
\usepackage{color}
\usepackage[normalem]{ulem}
 
\usepackage{url}
\usepackage[section]{placeins}
\usepackage[colorlinks=true,linkcolor=blue,citecolor=blue,urlcolor=blue]{hyperref}

\begin{document}


\title{Orientational Ordering of Janus Particle Arrays as Classical Compass Spin Systems}

\author{Zhijie Fan}
\email{zfanac@ustc.edu.cn}
\affiliation{Shanghai Research Center for Quantum Science and CAS Center for Excellence in Quantum Information and Quantum Physics, University of Science and Technology of China, Shanghai 201315, China }
\affiliation{Hefei National Laboratory, University of Science and Technology of China, Hefei 230088, China}

\author{Myeonggon Park}
\email{myeonggonpark@brandeis.edu}
\affiliation{Martin A. Fisher School of Physics, Brandeis University, Waltham, MA 02453, USA}

\author{Marija Vucelja}
\email{mv8h@virginia.edu}
\affiliation{Department of Physics, University of Virginia, Charlottesville, VA 22904, USA}
\affiliation{Department of Mathematics,  University of Virginia, Charlottesville, VA 22904, USA}

\author{Gia-Wei Chern}
\email{gc6u@virginia.edu}
\affiliation{Department of Physics, University of Virginia, Charlottesville, VA 22904, USA}

\date{\today}

\begin{abstract}
Janus-particle arrays provide a soft-matter platform in which particle orientations act as classical spins with anisotropic, bond-dependent interactions. Motivated by recent experiments on triangular arrays of metallodielectric Janus particles exhibiting sixfold orientational order and directly observable vortices and antivortices, we study effective models beyond the isotropic XY description. A classical $120^\circ$ compass model captures the minimal bond-directional interaction, while an effective Kern--Frenkel model incorporates the geometry of the Janus surface interaction more directly. Monte Carlo simulations show that both models exhibit a low-temperature sixfold ordered phase, a high-temperature disordered phase, and an intermediate quasi-long-range-ordered regime consistent with two Berezinskii--Kosterlitz--Thouless transitions. In both cases, the sixfold order arises through thermal order-by-disorder from a continuously degenerate ferromagnetic manifold. We further show that higher-order extensions of the $120^\circ$ interaction can alter the orientational selection: a quadratic term reverses the fluctuation-induced sixfold anisotropy, while a cubic term lifts the degeneracy energetically and selects the lattice-aligned directions that coincide with those observed experimentally.
These results connect the microscopic anisotropy of Janus-particle interactions to emergent XY-like vortex physics and illustrate how distinct microscopic mechanisms can produce similar sixfold orientational order.
\end{abstract}

\maketitle

\section{Introduction}

\label{sec:intro}

Colloidal systems have long provided versatile experimental platforms for exploring fundamental problems in statistical and condensed-matter physics. Their mesoscopic length and time scales allow individual particles to be directly resolved and tracked, enabling microscopic visualization of collective phenomena such as crystallization, melting, glass formation, and defect-mediated phase transitions \cite{murray1987,wang2012,deutschlander2015,thorneywork2017}. Advances in particle synthesis and external-field control have extended this paradigm from spherical particles with essentially isotropic interactions to anisotropic colloids with internal orientational degrees of freedom~\cite{hu2012,zhang2015,zhang2017,bianchi2017}. Janus particles, whose two hemispheres possess distinct physical or chemical properties, provide an especially flexible realization: their interactions can strongly depend on particle orientation and can be tuned through surface composition, solvent conditions, and external electric or magnetic fields~\cite{hu2012,sciortino2009,kumar2013,zhang2015,zhang2017}. In particular, metallodielectric Janus particles driven by AC electric fields exhibit tunable electrostatic and electrohydrodynamic interactions \cite{gangwal2008,yan2016,nishiguchi2018}.

\begin{figure*}
\centering
\includegraphics[width=1.85\columnwidth]{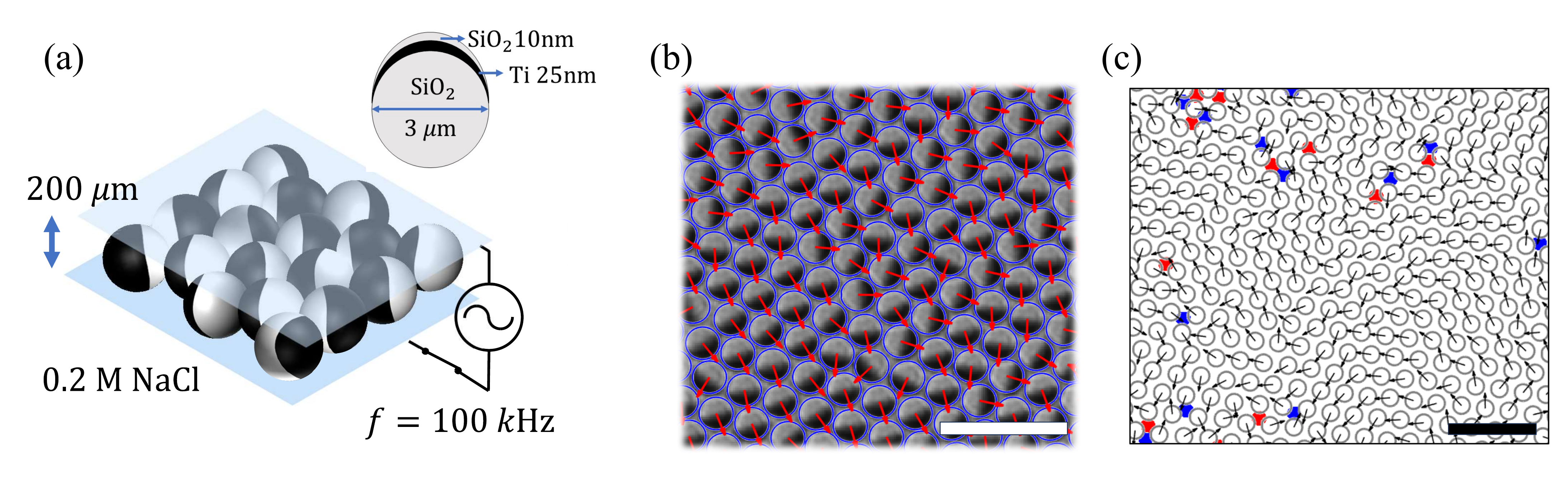}
\caption{\textbf{Experimental realization and effective-spin representation of the Janus-particle array.} (a) Schematic of the experimental setup, showing the metallic and dielectric hemispheres of a Janus particle and the two-dimensional colloidal monolayer subject to an AC electric field. (b) Representative region of the Janus-particle array, with the particle orientations represented by effective classical spins (red arrows). The scale bar is $10~\mu{\rm m}$. (c) Effective-spin representation of an orientational configuration, highlighting vortices (red) and antivortices (blue). The scale bar is $10~\mu{\rm m}$. The figure is reproduced from Ref.~\cite{park2025}.
\label{fig:exp}}
\end{figure*}

An interesting regime arises when Janus particles assemble into a crystalline array while retaining active orientational degrees of freedom~\cite{jiang2014,preisler2016,park2025}. Once translational motion is sufficiently suppressed, the particle orientations become the relevant low-energy degrees of freedom and can be represented by classical spins. This connects Janus-particle arrays to a broader class of colloidal systems whose orientational ordering can be described by effective spin Hamiltonians~\cite{sarlah2007,han2008,ortizambriz2016,libal2018}. Such systems combine the statistical mechanics of classical spin models with the experimental accessibility of colloids: individual effective spins can be directly imaged, interactions can be externally controlled, and topological defects can be tracked at the single-particle level. Recent experiments on triangular arrays of electrically polarized Janus particles provide a particularly interesting example~\cite{park2025}. As interaction strength increases, the particles develop strong orientational correlations, accompanied by readily observable vortices and antivortices [Fig.~\ref{fig:exp}]. The measured correlations and defect behavior exhibit signatures of Berezinskii--Kosterlitz--Thouless (BKT) physics, motivating comparison with the classical two-dimensional XY model ~\cite{park2025}.

However, the microscopic interactions between Janus particles differ fundamentally from the isotropic exchange interaction of the XY model. Dipolar and related electromagnetic interactions couple the particle orientations to the spatial direction connecting a pair of particles. The effective interaction therefore depends not only on the relative orientation of two ``spins,'' but also on their orientations relative to the underlying lattice bond. This coupling between spin and spatial degrees of freedom is the defining feature of compass-type interactions~\cite{nussinov2015}. In Janus-particle arrays, such bond dependence arises naturally from the anisotropic electromagnetic response of the particles rather than from an additional anisotropy imposed on an otherwise isotropic spin model. Closely related orientation-dependent interactions also underlie patchy-particle and Kern--Frenkel descriptions of Janus colloids~\cite{kern2003,sciortino2009,sciortino2010,giacometti2014,preisler2014,preisler2016,bianchi2017}.

Compass models form a broad class of anisotropic spin Hamiltonians, with prominent examples including the $120^\circ$ and Kitaev models~\cite{mostovoy2002,kitaev2006,nussinov2015}. Bond-dependent interactions arise naturally in effective descriptions of orbital degrees of freedom in transition-metal compounds~\cite{pen1997,mostovoy2002,nussinov2004,biskup2005}, frustrated orbital systems~\cite{chern2010}, and spin-orbit-coupled Mott insulators~\cite{jackeli2009,chaloupka2010}. Their classical counterparts also exhibit rich statistical behavior, including fluctuation-induced ordering and order-by-disorder phenomena~\cite{nussinov2004,mishra2004,biskup2005,wenzel2008}.

Janus-particle arrays therefore provide a soft-matter realization of the bond-directional interactions familiar from compass models in orbital and magnetic systems. Unlike their solid-state counterparts, however, the effective spins can be directly imaged at the single-particle level, allowing domains, fluctuations, and topological defects of compass-type spin systems to be observed in real space and real time. Janus arrays thus offer an experimentally accessible platform for exploring the statistical mechanics and dynamics of classical compass models. This viewpoint is particularly relevant to the recent Janus-lattice experiments: although the particle orientations define effective XY spins, the orientational distribution develops six preferred directions tied to the triangular lattice \cite{park2025}. Thus the degrees of freedom are XY-like, while the underlying interaction does not possess the continuous rotational symmetry of the XY Hamiltonian.

In this paper, we investigate the orientational ordering of triangular Janus-particle arrays using a hierarchy of effective models that incorporate the microscopic bond-dependent anisotropy. We first consider the classical $120^\circ$ model as a minimal realization of such coupling \cite{mostovoy2002,nussinov2004,biskup2005,nussinov2015}. Although the $120^\circ$ model has a long history in orbital and compass physics, the finite-temperature statistical mechanics of its classical triangular-lattice realization has, to our knowledge, not been systematically investigated. We find that thermal order-by-disorder lifts its accidental continuous ground-state degeneracy, producing sixfold orientational order and an intermediate quasi-long-range-ordered regime consistent with two BKT transitions. We then consider an effective Kern--Frenkel model that more directly represents the patch-like interaction between Janus particles \cite{kern2003,sciortino2010,giacometti2014,preisler2014,preisler2016}. Despite its very different microscopic form, the uniform ferromagnetic states remain continuously degenerate in energy, and thermal fluctuations again select six discrete orientations through order-by-disorder. 

Finally, we explore higher-order extensions of the $120^\circ$ interaction and show that nonlinear terms can qualitatively alter the mechanism and outcome of orientational selection. A quadratic correction preserves the continuous ground-state degeneracy but reverses the sign of the fluctuation-induced sixfold anisotropy beyond a critical coupling, thereby 
selects lattice-aligned directions that coincide with those observed experimentally, 
whereas a cubic correction lifts the degeneracy directly and selects the same orientations energetically. Together, these results show how distinct microscopic interactions and degeneracy-lifting mechanisms can give rise to closely related sixfold ordering and XY-like vortex physics in triangular Janus arrays \cite{park2025}.

The rest of the paper is organized as follows. In Sec.~II, we review the experimental Janus-particle system and discuss the physical origin and symmetry of its effective orientational interactions. In Sec.~III, we introduce the triangular-lattice $120^\circ$ model and present Monte Carlo results for its finite-temperature ordering and topological defects. In Sec.~IV, we formulate an effective Kern--Frenkel model for the Janus-particle interaction and analyze its fluctuation-induced orientational ordering and thermodynamic behavior. In Sec.~V, we investigate higher-order extensions of the $120^\circ$ interaction and compare fluctuation-induced and energetic mechanisms of sixfold orientational selection. Finally, Sec.~VI presents the discussion and outlook.

\section{Janus-Particle Array and Effective Interactions}

The system motivating our study is the two-dimensional Janus colloidal crystal investigated experimentally in Ref.~\cite{park2025}. The Janus particles consist of silica spheres of diameter $3~\mu{\rm m}$ coated with a thin titanium layer on one hemisphere, producing metallic and dielectric sides [Fig.~\ref{fig:exp}(a)]. The particles are suspended in an aqueous electrolyte solution and sediment under gravity to form a two-dimensional monolayer. At sufficiently high density, their positions develop crystalline order on a triangular lattice while the individual particles retain rotational freedom. The particle centers therefore form an approximately fixed lattice, leaving the orientations as the relevant fluctuating degrees of freedom.

\begin{figure}
\centering
\includegraphics[width=0.9\columnwidth]{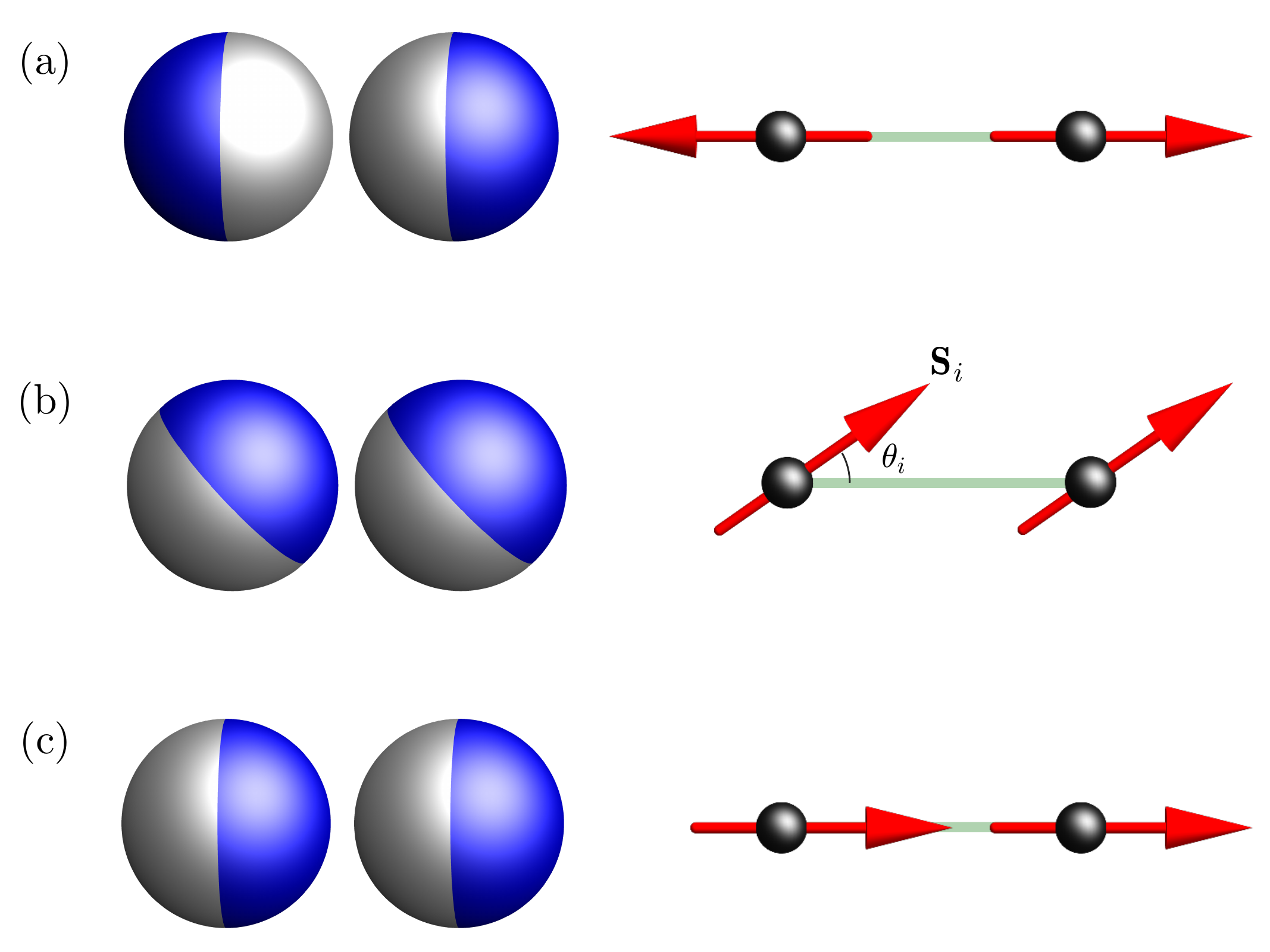}
\caption{\textbf{Mapping of Janus-particle orientations onto effective XY spins and the origin of bond-dependent interactions.} The AC electric field along the $z$ direction constrains the normal to the interface between the metallic and dielectric hemispheres to the $xy$ plane, leaving a single azimuthal angle $\theta$ represented by an effective spin $\mathbf{S}$, pointing from the dielectric hemisphere (white) to the metal patch (blue) direction. For a neighboring pair, the antiparallel configuration in (a) is energetically unfavorable, while the parallel configurations in (b) and (c) are inequivalent because their energies also depend on the spin orientations relative to the interparticle bond. The bond-aligned configuration in (c) is favored, motivating the ferromagnetic $120^\circ$ interaction introduced in Sec.~III.
\label{fig:pair}
}
\end{figure}

An AC electric field is applied perpendicular to the colloidal monolayer, along the $z$ direction [Fig.~\ref{fig:exp}(a)]. The experiments in Ref.~\cite{park2025} were performed at a fixed AC frequency of 100 kHz, whereas the induced interparticle interactions generally depend on the driving frequency. In addition to inducing interactions between the particles, the field effectively constrains their orientations. The particles preferentially orient with the interface separating the metallic and dielectric hemispheres along the $z$ direction, so that the normal to this interface lies in the $xy$ plane. The remaining orientational degree of freedom is therefore described, to a good approximation, by an effective spin \begin{equation}
    \mathbf{S}_i=(\cos\theta_i,\sin\theta_i).
    \label{eq:xyspin}
\end{equation}
defined to point from the dielectric to the metallic hemisphere, with $\theta_i$ measured relative to the horizontal $x$-axis (a principal triangular-lattice vector), as illustrated in Fig.~\ref{fig:pair}.

Increasing the applied field strengthens the interactions between neighboring particles and drives the orientational degrees of freedom toward collective order. The experiments reveal smoothly varying spin textures containing vortices and antivortices with winding numbers $+1$ and $-1$, respectively [Fig.~\ref{fig:exp}(b,c)]. The orientational correlations, susceptibility, and vortex density exhibit signatures of Berezinskii--Kosterlitz--Thouless (BKT) physics, motivating comparison with the two-dimensional XY model \cite{park2025}. We denote the isotropic nearest-neighbor XY Hamiltonian by
\begin{equation}
	\mathcal H_{\rm XY} = J_{\rm XY}\sum_{\langle ij\rangle} \mathbf S_i\cdot\mathbf S_j,
\label{eq:xy-model}
\end{equation}
where the interaction depends only on the relative orientation of neighboring spins. With this sign convention, $J_{\rm XY}<0$ corresponds to the conventional ferromagnetic XY model. At stronger coupling, however, the orientational distribution of the Janus array develops six distinct peaks associated with the triangular-lattice symmetry. Thus, while the orientational degrees of freedom are well represented by XY spins, their microscopic interaction cannot be described by the isotropic Hamiltonian in Eq.~\eqref{eq:xy-model} alone.

The origin of this additional anisotropy lies in the polarization of the metallodielectric Janus particles. Under the applied electric field, the metallic and dielectric hemispheres have different dielectric responses and develop distinct induced-charge distributions \cite{gangwal2008,yan2016,nishiguchi2018}. The resulting interaction depends not only on the relative orientation of neighboring particles but also on their orientations relative to the bond connecting them. For the planar degrees of freedom considered here, this bond dependence can be represented schematically by
\begin{equation}
	V_{ij} = \mathcal V \left( \mathbf S_i\cdot\hat{\mathbf r}_{ij}, \mathbf S_j\cdot\hat{\mathbf r}_{ij} \right),
\label{eq:general-interaction}
\end{equation}
where $\hat{\mathbf r}_{ij}$ denotes the bond direction. Such bond-directional terms distinguish the effective Janus-particle interaction from the isotropic XY model and provide the microscopic motivation for the compass-type Hamiltonians considered below.

The coupling between particle orientation and bond direction is the essential structure of compass-type interactions \cite{nussinov2015}. On the triangular lattice, the three symmetry-related nearest-neighbor bond directions naturally generate such bond-dependent coupling and ultimately allow the orientational order to lock to the underlying lattice. The detailed interaction between metallodielectric Janus particles can involve nonuniform induced charge, electrolyte screening, and electrohydrodynamic effects, and we do not attempt here to derive a microscopic pair potential from the experimental parameters. Instead, we consider a hierarchy of complementary effective descriptions: the $120^\circ$ model as a minimal compass-type Hamiltonian on the triangular lattice, an effective Kern--Frenkel model that more directly represents the orientation-dependent interaction between Janus particles, and higher-order extensions of the $120^\circ$ interaction that allow us to explore different mechanisms of orientational selection.

\section{Triangular-Lattice $120^\circ$ Model}

The effective interaction between neighboring Janus particles can be understood qualitatively from their polarization under the applied AC electric field. The distinct dielectric responses of the metallic and insulating hemispheres produce asymmetric, oscillating polarization and charge distributions, with a particularly strong response on the metallic side \cite{gangwal2008,yan2016,nishiguchi2018}. Configurations that bring the strongly polarized metallic hemispheres into close proximity are therefore energetically unfavorable. This disfavors the antiparallel configuration in Fig.~\ref{fig:pair}(a), giving the effective orientational interaction an overall ferromagnetic character.
The interaction is nevertheless strongly orientation-dependent. Although both Figs.~\ref{fig:pair}(b) and \ref{fig:pair}(c) represent parallel spins, the bond-aligned configuration in Fig.~\ref{fig:pair}(c) better separates the metallic hemispheres and is energetically favored.

\begin{figure*}
\centering
\includegraphics[width=1.99\columnwidth]{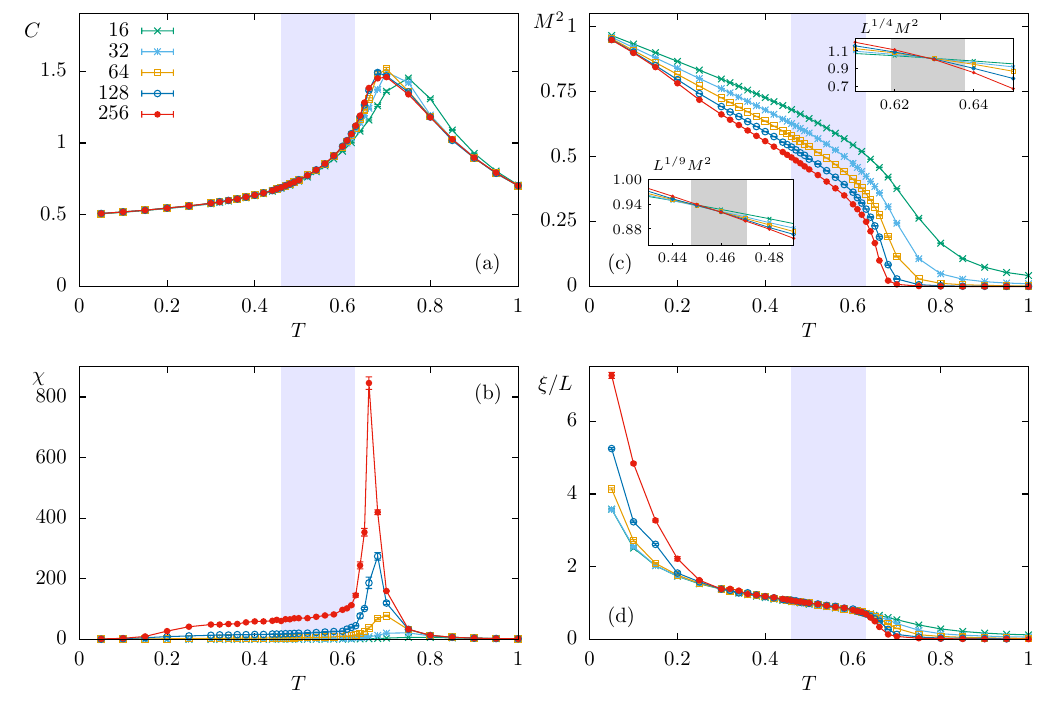}
\caption{\textbf{Monte Carlo results for the triangular-lattice $120^\circ$ model for system sizes up to $L=256$.} Temperature dependence of (a) the specific heat $C$, (b) the magnetic susceptibility $\chi$, (c) the squared magnetization $M^2$, and (d) the correlation-length ratio $\xi/L$. 
The crossings for the two largest sizes give rough estimates of the lower and upper BKT temperatures, $T_{\mathrm{BKT}}^{\mathrm{lower}}/J\simeq0.46$ and $T_{\mathrm{BKT}}^{\mathrm{upper}}/J\simeq0.63$.
The purple shading indicates the estimated intermediate quasi-long-range-ordered region, while the gray bands in the insets mark approximate crossing ranges.
Finite-size corrections, especially near the lower transition, may further shift the transition temperatures.
}
\label{fig:120mc}
\end{figure*}

To isolate the bond-directional component, we consider the minimal pure compass bilinear interaction, the $120^\circ$ model,
\begin{equation}
	\mathcal{H}_{\rm 120^\circ} = -J\sum_{\langle ij\rangle} \left(\mathbf S_i\cdot\hat{\mathbf r}_{ij}\right) \left(\mathbf S_j\cdot\hat{\mathbf r}_{ij}\right),
\label{eq:120model}
\end{equation}
where $J>0$, $\mathbf S_i$ is the effective XY spin defined in Eq.~(\ref{eq:xyspin}), and $\hat{\mathbf r}_{ij}$ is the unit vector along the nearest-neighbor bond connecting sites $i$ and $j$. It is useful to rewrite Eq.~\eqref{eq:120model} as
\begin{equation}
	\mathcal{H}_{120^\circ} = -\frac{3NJ}{2} + \frac{J}{2}\sum_{\langle ij\rangle} \left[\left(\mathbf S_i - \mathbf S_j\right)\cdot\hat{\mathbf r}_{ij}\right]^2,
\label{eq:120model-rewrite}
\end{equation}
where we have used the angle-independent identity $\sum_{j\in{\rm n.n.}(i)}(\mathbf S_i\cdot\hat{\mathbf r}_{ij})^2=2\sum_{\gamma=1}^3(\mathbf S_i\cdot\hat{\mathbf e}_\gamma)^2=3$ for a unit XY spin on the triangular lattice. This form makes explicit the lower bound $\mathcal H\geq -3NJ/2$.

For an isolated bond, the interaction is minimized by parallel spins aligned with the bond. On the triangular lattice, the three symmetry-related bond families can be represented by projection vectors separated by $120^\circ$, giving the model its characteristic compass structure. The $120^\circ$ model was originally introduced in the context of orbital ordering in transition-metal compounds, including triangular-lattice nickelates such as NaNiO$_2$ \cite{mostovoy2002}, and has since become a prototypical compass model \cite{nussinov2004,biskup2005,nussinov2015}. Related interactions also arise in frustrated orbital systems on the pyrochlore lattice \cite{chern2010}. Its classical finite-temperature statistical mechanics on the triangular lattice, however, remains largely unexplored.

Interestingly, despite its explicit bond anisotropy, the model has an accidental continuous degeneracy within its ferromagnetic ground-state manifold. For any uniform configuration $\mathbf S_i=\mathbf S=(\cos\Theta,\sin\Theta)$, the second term in Eq.~\eqref{eq:120model-rewrite} vanishes, so the lower bound is attained with $E_g=-3NJ/2$, independent of the common orientation angle $\Theta$. All uniformly polarized states are therefore degenerate, even though global $SO(2)$ rotation is not a symmetry of the Hamiltonian. This accidental degeneracy provides a natural setting for fluctuation-induced order \cite{mostovoy2002,nussinov2004,biskup2005,nussinov2015}.

The pure $120^\circ$ model in Eq.~\eqref{eq:120model} should be viewed as the minimal bond-directional component of a more general interaction. At the bilinear level, a symmetry-allowed nearest-neighbor Hamiltonian may contain both the isotropic XY interaction defined in Eq.~\eqref{eq:xy-model} and the compass term in Eq.~\eqref{eq:120model}. In fact, as shown in Appendix~\ref{app:effective-interaction}, a simple multipolar description of the AC-induced polarization of the Janus particles naturally generates both contributions: an antiferromagnetic isotropic exchange with $J_{\rm XY}>0$ and a stronger ferromagnetic $120^\circ$ compass interaction, together with higher-order bond-dependent terms. In the present section, we set $J_{\rm XY}=0$ and focus on the pure compass model in order to isolate the effects of bond-directional anisotropy and its associated order-by-disorder physics.

We investigate the finite-temperature behavior of Eq.~(\ref{eq:120model}) using classical Monte Carlo simulations on $L\times L$ triangular lattices with periodic boundary conditions. Spins are updated sequentially using a standard local Metropolis algorithm, with each trial move corresponding to a random rotation of a single spin, supplemented by collective global spin rotations. The range of single-spin trial rotations is adjusted during thermalization to maintain an acceptance rate of approximately $50\%$. We consider system sizes up to $L=256$ and temperatures ranging from $T=0.05$ to $1.00$, with additional temperature points near the transition region. Measurements are taken every $20$ Monte Carlo sweeps after thermalization, with approximately $2\times10^5$–$2\times10^6$ production measurements accumulated per temperature and system size. Statistical uncertainties are estimated using the binning and jackknife method.

Figure~\ref{fig:120mc} summarizes the thermal evolution of the magnetic state through the specific heat, magnetization, susceptibility, and correlation length. The specific heat,
\begin{align}
C = \frac{\langle E^2 \rangle - \langle E\rangle^2}{N T^2}\,,
\end{align}
exhibits a broad maximum near $T/J\sim0.7$, signaling the development of substantial magnetic correlations upon cooling rather than a sharp thermodynamic transition. This crossover is accompanied by pronounced finite-size effects in the magnetic observables. In particular, the squared magnetization,
\begin{align}
M^2 = \left|\frac{1}{N}\sum_i \mathbf S_i\right|^2\,,
\end{align}
decreases rapidly with increasing system size at high temperatures, as expected for a disordered phase with only short-range correlations. At sufficiently low temperatures, by contrast, $M^2$ approaches a nonzero value with increasing $L$, consistent with the development of long-range ferromagnetic order. The corresponding magnetic susceptibility,
\begin{align}
\chi = \frac{N}{T}\left(\langle M^2\rangle - \langle |M| \rangle^2\right)\,,
\end{align}
is strongly enhanced in the intervening temperature range, further indicating large magnetic fluctuations.

The nature of this intermediate regime is seen more clearly from the second-moment correlation-length ratio $\xi/L$. We compute
\begin{align}
\xi = \frac{1}{2\sin(k_{\min}/2)}\sqrt{\frac{\langle M^2 \rangle}{\langle M^2(\mathbf{k}_{\min})\rangle}-1}\,,
\end{align}
where $M^2(\mathbf{k}) = \left|\frac{1}{N}\sum_i \mathbf{S}_i e^{-i\mathbf{k}\cdot\mathbf{r}_i}\right|^2$~\cite{caracciolo1993,edwards1989}. The estimate is averaged over the two symmetry-equivalent smallest nonzero wavevectors $\mathbf{k}_{\min}=(0,4\pi/\sqrt{3}L)$ and $(2\pi/L,-2\pi/\sqrt{3}L)$. Rather than exhibiting a single crossing characteristic of an isolated continuous transition, the $\xi/L$ curves for different system sizes approximately merge over an extended temperature interval. Together with the persistent size dependence of $M^2$, this behavior is consistent with an intermediate critical phase characterized by quasi-long-range magnetic order~\cite{viet2009,ding2014,tuan2022,yao2025nonclassicalregimetwodimensionallongrange__universality,xiao_two-dimensional_2024__universality}.

\begin{figure}
\centering
\includegraphics[width=0.98\columnwidth]{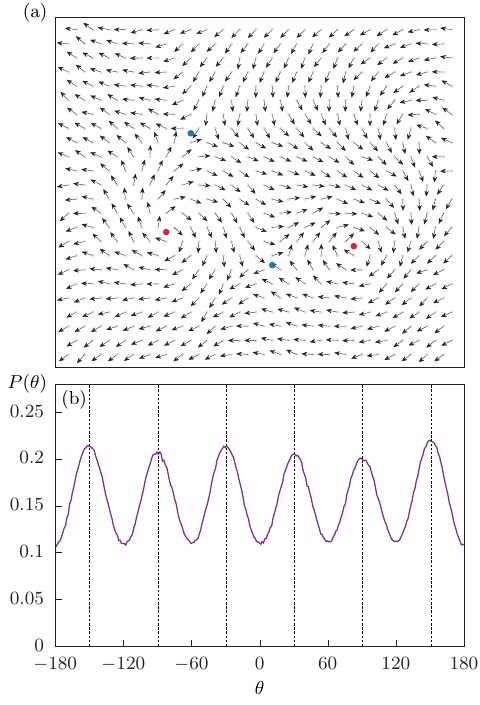}
\caption{\textbf{Low-temperature configurations and angular distribution of the triangular-lattice $120^\circ$ model at $L=24$ and $T=0.05J$.} 
(a) Representative spin configuration showing surviving vortex--antivortex pairs during the relaxation, with the red dots marking vortices and the blue dots marking anti-vortices. The snapshot illustrates the topological character of the defects and is not an equilibrium vortex-statistics measurement. (b) Distribution of individual spin angles, $P(\theta)$, accumulated over all lattice sites in $10^4$ independently initialized runs after equilibration. The distribution shows six preferred orientations near $\theta=\pi/6+n\pi/3$, $n=0,\ldots,5$, marked by the dashed lines.}
\label{fig:120snapshot}
\end{figure}

This phase structure is characteristic of the two-BKT-transition scenario of the two-dimensional six-state clock model and, more generally, XY models with sixfold anisotropy \cite{kosterlitz1973,kosterlitz1974,jose1977,tobochnik1982,challa1986,tomita2002}. In the intermediate critical phase, the sixfold anisotropy is irrelevant, and the spin correlations decay algebraically,
\begin{equation}
    \left\langle \mathbf S(\mathbf 0)\cdot\mathbf S(\mathbf r)\right\rangle
    \sim r^{-\eta(T)},
    \label{eq:powerlaw}
\end{equation}
with a continuously varying exponent $\eta(T)$. The upper BKT transition separates this critical phase from the high-temperature disordered phase via vortex--antivortex unbinding, whereas at the lower transition, the sixfold anisotropy becomes relevant and locks the system into one of six symmetry-related ordered states \cite{jose1977}. The corresponding anomalous exponents are $\eta=1/4$ and $\eta=1/9$ at the upper and lower transitions, respectively. Similar two-stage BKT ordering occurs in other frustrated spin systems with an effective sixfold order parameter \cite{chern2012}.

The finite-size behavior of our Monte Carlo data is consistent with this scenario. Within the critical phase, $M^2\sim L^{-\eta(T)}$, so that $M^2L^\eta$ is approximately size-independent when the corresponding value of $\eta$ is realized. Using the expected values $\eta=1/4$ and $1/9$, a rough finite-size estimation gives $T_{\rm BKT}^{\rm upper}/J\simeq0.63$ and $T_{\rm BKT}^{\rm lower}/J\simeq0.46$, respectively. Corrections to scaling, especially near the lower BKT transition, may further shift the thermodynamic-limit values. The behavior of $\xi/L$ also identifies approximately the same intermediate regime, with curves for different $L$ merging between the two transition temperatures. The simulations therefore support a high-temperature disordered phase, an intermediate BKT phase with quasi-long-range order, and a low-temperature ferromagnetic phase with sixfold orientational symmetry breaking.

The microscopic spin configurations provide a complementary view of this behavior. Figure~\ref{fig:120snapshot}(a) shows a representative configuration during Monte Carlo relaxation at $T=0.01\,J$. 

Vortices and antivortices remain clearly identifiable in the effective XY-spin texture despite the explicitly bond-dependent interaction of the \(120^\circ\) model. At this low temperature, the defects occur predominantly as bound vortex--antivortex pairs, consistent with the suppression of free vortices below the upper BKT transition. However, because this snapshot is taken during relaxation rather than from the equilibrated ensemble, the surviving vortex––antivortex pairs should be viewed as an illustration of the topological defects supported by the model, rather than as an equilibrium measure of vortex statistics.

The sixfold orientational preference is visible in the angular distribution $P(\theta)$ in Fig.~\ref{fig:120snapshot}(b). Here, $\theta_i$ is the angle of an individual spin, $\mathbf{S}_i=(\cos\theta_i,\sin\theta_i)$, measured relative to a principal symmetry axis of the triangular lattice. The distribution is obtained by pooling local spin angles over lattice sites and sampled configurations. It exhibits six peaks near $\theta=\pi/6+n\pi/3$, with $n=0,\ldots,5$. For a uniform classical ground state, all spins share a common angle $\Theta$, and the energy is independent of $\Theta$. The observed preference for six orientations is therefore consistent with thermal order-by-disorder, in which thermal fluctuations lift the accidental continuous degeneracy. Interestingly, quantum fluctuations in the corresponding orbital $120^\circ$ model select a different set of six states, with orientations $\Theta=n\pi/3$ \cite{mostovoy2002}. Thermal and quantum order-by-disorder thus favor two interlaced sets of orientations, shifted from one another by $\pi/6$ within the same classically degenerate manifold.


\section{Effective Kern--Frenkel Model}

\begin{figure*}
\centering
\includegraphics[width=1.99\columnwidth]{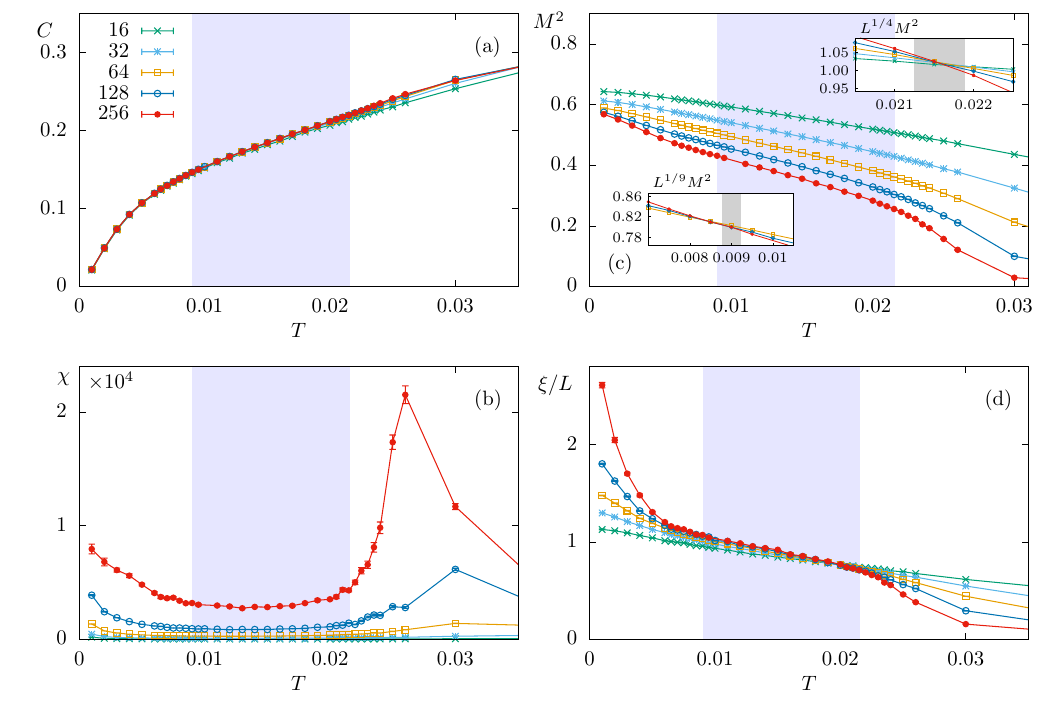}
  \caption{\textbf{Monte Carlo results for the effective Kern--Frenkel model for system sizes up to $L=256$.}
  Temperature dependence of (a) the specific heat $C$, (b) the magnetic susceptibility $\chi$, (c) the squared magnetization $M^2$, and (d) the correlation-length ratio $\xi/L$.
  The lower-left and upper-right insets in panel (c) show $L^{1/9}M^2$ and $L^{1/4}M^2$, respectively, following the same scaling analysis as for the $120^\circ$ model.
  The crossings for the two largest sizes give rough estimates of the lower and upper BKT temperatures, $T_{\mathrm{BKT}}^{\mathrm{lower}}/J\simeq0.009$ and $T_{\mathrm{BKT}}^{\mathrm{upper}}/J\simeq0.02$.
  The purple shading indicates the estimated intermediate quasi-long-range-ordered region, while the gray bands in the insets mark approximate crossing ranges.}
\label{fig:kfmc}
\end{figure*}

The $120^\circ$ model provides a minimal bilinear description of the bond-dependent interaction, but the interaction between Janus particles need not be well represented by a quadratic form. A complementary description is motivated by the Kern--Frenkel (KF) framework, originally introduced for particles with localized surface patches \cite{kern2003}. KF-type interactions have since been widely used to describe the self-assembly and phase behavior of patchy and Janus particles \cite{sciortino2009,sciortino2010,giacometti2014,preisler2014,preisler2016}. Their defining feature is an explicit dependence on the orientation of a surface patch relative to the line connecting a pair of particles, making this framework naturally suited to the anisotropic interactions considered here.

We construct an effective KF-type interaction for the repulsive metallic hemispheres of the Janus particles. Since the particle orientations are restricted to the $xy$ plane, the orientation of the metallic hemisphere is specified by the same effective spin $\mathbf S_i$ introduced in Eq.~(\ref{eq:xyspin}). For a neighboring pair $i$ and $j$, whether the metallic region of particle $i$ faces particle $j$ is determined by the projection of $\mathbf S_i$ onto the bond direction $\hat{\mathbf r}_{ij}$. We introduce the angular function
\begin{equation}
f(x;\theta_0)=
\begin{cases}
1, & x>\cos\theta_0,\\
0, & \mathrm{otherwise},
\end{cases}
\label{eq:kfpatch}
\end{equation}
where $\theta_0$ specifies the angular extent of the metallic patch. Here, we set $\cos\theta_0=0.01$, corresponding to $\theta_0$ slightly smaller than $\pi/2$ and hence to an approximately hemispherical metallic patch. The factors $f(\mathbf S_i\cdot\hat{\mathbf r}_{ij};\theta_0)$ and $f(\mathbf S_j\cdot\hat{\mathbf r}_{ji};\theta_0)$ impose the angular conditions for the metallic regions of the two particles to face one another.

To incorporate the orientation dependence of the repulsive interaction within this patch-based construction, we consider the effective Hamiltonian
\begin{equation}
	\mathcal H_{\rm KF} = J\sum_{\langle ij\rangle} f(\mathbf S_i\cdot\hat{\mathbf r}_{ij};\theta_0) f(\mathbf S_j\cdot\hat{\mathbf r}_{ji};\theta_0) 
	\left(\mathbf S_i\cdot\hat{\mathbf r}_{ij}\right) \left(\mathbf S_j\cdot\hat{\mathbf r}_{ji}\right),
\label{eq:kfmodel}
\end{equation}
with $J>0$. The two $f$ factors impose the geometrical patch conditions, while the bond projections introduce a continuous dependence of the repulsive interaction strength on particle orientation. When the patch conditions are not simultaneously satisfied, the interaction is zero. Thus, Eq.~(\ref{eq:kfmodel}) retains only the repulsive interaction associated with facing metallic regions and could be viewed as an effective KF-type interaction adapted to the orientational degrees of freedom of the present Janus array.

For the choice $\cos\theta_0=0.01$, the KF model exhibits an important yet somewhat unexpected degeneracy. Consider a spatially uniform ferromagnetic configuration, $\mathbf S_i=\mathbf S$ for all $i$. For any bond, the two patch conditions would require simultaneously
\begin{equation}
\mathbf S\cdot\hat{\mathbf r}_{ij}>\cos\theta_0,
\qquad
-\mathbf S\cdot\hat{\mathbf r}_{ij}>\cos\theta_0.
\end{equation}
Since $\cos\theta_0>0$, these two inequalities cannot be satisfied simultaneously; consequently, the metallic patches never face one another in a perfectly ferromagnetic configuration, and every bond contributes zero interaction energy. All uniform ferromagnetic states are therefore exactly degenerate, independent of the global orientation of the spins.

The sixfold orientational selection observed at low temperature thus arises from thermal fluctuations about the continuously degenerate ferromagnetic manifold. Departures from perfect alignment can bring neighboring metallic patches into contact and activate repulsive interactions, and the available spectrum of such fluctuations depends on the global orientation of the ordered state relative to the triangular lattice. Thermal fluctuations therefore lift the degeneracy entropically and select a discrete set of orientations. The effective KF model consequently provides a distinct realization of thermal order-by-disorder in a patchy-particle system, despite its microscopic interaction being very different from the bilinear $120^\circ$ model.

\begin{figure}
\centering
\includegraphics[width=0.98\columnwidth]{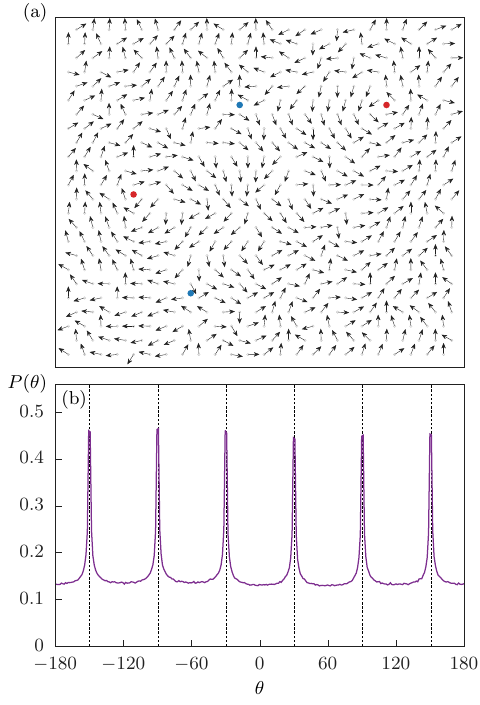}
\caption{\textbf{Low-temperature configurations and angular distribution of the effective Kern--Frenkel model at $L=24$ and $T=0.01J$.} 
(a) Representative spin configuration showing orientational textures and surviving vortex--antivortex pairs during relaxation. The snapshot illustrates the defect structure and is not a measurement of equilibrium vortex statistics. Compared with the $120^\circ$ model, the local spin texture is noticeably less smooth, consistent with the patch-dependent nature of the KF interaction.
(b) Distribution of individual spin angles, $P(\theta)$, accumulated over $10^4$ independently initialized run's after equilibration. Here, $\theta$ is the local spin angle measured relative to a principal symmetry axis of the triangular lattice. The pronounced peaks near $\theta=\pi/6+n\pi/3$, $n=0,\ldots,5$, indicate fluctuation-induced selection of the same six preferred orientations as in the $120^\circ$ model.}
\label{fig:kfsnapshot}
\end{figure}


We study Eq.(\ref{eq:kfmodel}) using the same Metropolis Monte Carlo approach as in the $120^\circ$ model. Measurements are taken every $20$ Monte Carlo sweeps after thermalization, accumulating approximately $5\times10^5$ measurements per temperature for $L=16$ and at least $4\times10^6$ for larger sizes. Independent runs are combined using inverse-variance weights, with statistical uncertainties estimated through binning and jackknife analysis. Figure\ref{fig:kfmc} shows the specific heat $C$, magnetic susceptibility $\chi$, squared magnetization $M^2$, and correlation-length ratio $\xi/L$ for $L=16$, $32$, $64$, $128$, and $256$. Despite the substantially different microscopic form of the interaction, its thermodynamic behavior closely resembles that of the $120^\circ$ model. The high-temperature phase is disordered, while the low-temperature behavior of $M^2$ is consistent with ferromagnetic long-range order. Between these regimes, the size dependence of $M^2$ and the approximate merging of $\xi/L$ over an extended temperature interval suggest quasi-long-range order. 

This behavior is consistent with the two-BKT-transition scenario of an XY system subject to an emergent sixfold anisotropy. As reviewed in Sec.~III, a sixfold perturbation permits two BKT transitions separated by a critical phase with algebraically decaying correlations \cite{jose1977,tobochnik1982,challa1986,tomita2002}. The upper transition is associated with vortex--antivortex unbinding and the loss of quasi-long-range order, while the lower transition marks the onset of true long-range order as the fluctuation-generated sixfold anisotropy becomes relevant. The shaded region in Fig.~\ref{fig:kfmc} indicates the intermediate BKT regime inferred from the finite-size behavior of $M^2$ and $\xi/L$. The two estimated BKT temperature is approximately $T_{\rm BKT}^{\rm upper}/J \simeq 0.02$ and $T_{\rm BKT}^{\rm lower}/J \simeq 0.009$. Thus, although the $120^\circ$ and KF models have very different microscopic interactions, both possess a continuously degenerate manifold of uniform ferromagnetic states whose degeneracy is lifted by thermal fluctuations, leading to closely related two-stage ordering behavior.

The microscopic configurations nevertheless reveal an important difference between the two models. In the KF model, neighboring spins interact only when fluctuations bring their metallic patches into mutual overlap. The orientational interaction is therefore highly intermittent: for many local configurations, and in particular for an exactly uniform ferromagnetic state, the corresponding bond interaction simply vanishes. As a result, the energetic penalty for gradual spatial variations of the spin orientation is comparatively weak, and the low-temperature configurations are noticeably less smooth than those of the $120^\circ$ model. The snapshot in Fig.~\ref{fig:kfsnapshot}(a) accordingly contains substantial local angular variation even within broadly ordered regions, together with clearly identifiable vortex--antivortex textures. As in Fig.~\ref{fig:120snapshot}(a), this relaxation snapshot is included to illustrate the vortex structure supported by the model; it is not used to infer equilibrium vortex populations or locate the BKT transitions.


At the same time, the distribution of individual spin angles, $P(\theta)$, in Fig.~\ref{fig:kfsnapshot}(b) displays a pronounced sixfold structure. The peaks occur near $\theta=\pi/6+n\pi/3$, $n=0,\ldots,5$, matching the orientations selected by thermal order-by-disorder in the $120^\circ$ model. This angular preference coexists with a relatively rough local spin texture: individual configurations can exhibit substantial spatial variations, while the distribution accumulated over lattice sites and equilibrated runs favors six preferred directions. The two models therefore exhibit the same sixfold order-by-disorder selection and similar finite-temperature BKT phenomenology, despite their different microscopic interactions and ordering temperature scales.

\section{Orientational selection from higher-order interactions}

The bilinear $120^\circ$ model represents the leading bond-dependent coupling between neighboring spins, but more general smooth interactions can contain higher powers of the same symmetry-allowed bond variable.
In fact, a multipole expansion of the two-center polarization description of Janus particles under an AC field can further generate quadratic and cubic powers of the bond variable in the locally parallel sector, as derived in Appendix~\ref{app:effective-interaction}.
To explore how such nonlinear corrections affect the orientational selection, we consider the generalized Hamiltonian
\begin{equation}
	\mathcal H_{\rm eff} = -\sum_{\langle ij\rangle} \left[ J_1 q_{ij} + J_2 q_{ij}^2 + J_3 q_{ij}^3 +\cdots \right],
\label{eq:higher120}
\end{equation}
where
\begin{equation}
q_{ij}
=
(\mathbf S_i\cdot\hat{\mathbf r}_{ij})
(\mathbf S_j\cdot\hat{\mathbf r}_{ij}).
\end{equation}
The $J_1$ term reduces to the original $120^\circ$ interaction, while $J_2$ and $J_3$ represent the leading nonlinear extensions constructed from the same bond-dependent variable $q_{ij}$. Although these higher-order terms preserve the symmetries of the underlying triangular lattice, they affect the continuously degenerate ferromagnetic manifold in qualitatively different ways.

For a spatially uniform ferromagnetic state, $\mathbf S_i=\mathbf S=(\cos\Theta,\sin\Theta)$, the bond variable becomes $q_{ij}=\cos^2(\Theta-\phi_\gamma)$, where $\phi_\gamma$ denotes one of the three bond directions of the triangular lattice. Summing over the three bond orientations gives $\sum_{\gamma=1}^{3}\cos^2(\Theta-\phi_\gamma)=3/2$ and $\sum_{\gamma=1}^{3}\cos^4(\Theta-\phi_\gamma)=9/8$, both independent of $\Theta$. Consequently, for $J_3=0$, the energy per site is
\begin{equation}
\frac{E_{\rm FM}}{N}=-\frac{3J_1}{2}-\frac{9J_2}{8},
\label{eq:j1j2fmenergy}
\end{equation}
and the continuous angular degeneracy of the ferromagnetic ground states remains exact even in the presence of the quadratic $J_2$ term. Any discrete orientational selection in the $J_1$--$J_2$ model must therefore arise from fluctuations rather than from the classical ground-state energy.

\begin{figure}[t]
\centering
\includegraphics[width=0.98\linewidth]{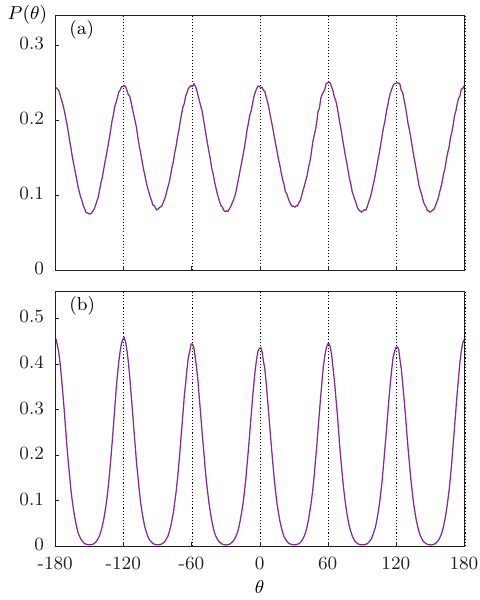}
  \caption{\textbf{Orientational selection in higher-order
  $120^\circ$ models at $L=24$ and $T=0.05J_1$.}
  The distributions $P(\theta)$ are accumulated over individual
  spin angles at all lattice sites in $10^4$ independently
  initialized runs after equilibration. Here, $\theta$ is measured
  relative to a principal lattice axis.
  (a) $J_1=1$, $J_2=0.1$, and $J_3=0$.
  The $J_2$ term preserves the degeneracy of uniform ferromagnetic
  states but modifies their fluctuation spectrum, shifting the
  thermally preferred orientations to $\theta=n\pi/3$,
  $n=0,\ldots,5$, relative to the pure $120^\circ$ model.
  (b) $J_1=1$, $J_3=0.1$, and $J_2=0$.
  The cubic term explicitly lifts the continuous degeneracy
  and energetically selects the same six orientations.
  The sharper peaks and lower probability between them indicate
  stronger orientational locking.}
\label{fig:higher_term_model_angle}
\end{figure}

The $J_2$ interaction nevertheless modifies the spectrum of fluctuations about different members of this degenerate manifold. As derived in Appendix~\ref{app:harmonic-selection}, a harmonic expansion around a uniform state produces an orientation-dependent fluctuation free energy, which may be expressed to leading angular order as
\begin{equation}
F_{\rm fl}(\Theta)=F_0+K_6(T,J_2/J_1)\cos6\Theta+\cdots.
\end{equation}
For the pure $J_1$ model, $K_6>0$, so thermal order-by-disorder favors $\cos6\Theta=-1$, corresponding to the six orientations
\begin{equation}
\Theta=\frac{\pi}{6}+n\frac{\pi}{3},
\qquad n=0,\ldots,5.
\end{equation}
The harmonic calculation shows, however, that increasing $J_2$ changes the sign of this fluctuation-generated sixfold anisotropy. The sign reversal occurs at approximately
\begin{equation}
\frac{J_2}{J_1}\simeq0.032.
\end{equation}
Above this value, $K_6<0$, and thermal fluctuations instead favor
\begin{equation}
\Theta=n\frac{\pi}{3},
\qquad n=0,\ldots,5,
\end{equation}
namely the six directions aligned with the principal axes of the triangular lattice. Thus, the nonlinear $J_2$ interaction provides an unusual example in which a higher-order term does not lift the classical ground-state degeneracy at all, but reverses the sign of the anisotropy generated by thermal order-by-disorder.

This behavior is illustrated in Fig.~\ref{fig:higher_term_model_angle}(a) for $J_1=1$, $J_2=0.1$, and $J_3=0$ at $T=0.05J_1$. Since $J_2/J_1=0.1$ lies well above the harmonic sign-reversal value, the angular distribution develops six maxima at $\Theta=n\pi/3$. These directions differ by $\pi/6$ from those selected in the pure $120^\circ$ model and, notably, coincide with the lattice-aligned orientations observed experimentally~\cite{park2025}. The selection remains entirely entropic: all uniform ferromagnetic configurations have exactly the same classical energy, but the lattice-aligned states possess the larger fluctuation entropy.

The cubic $J_3$ term behaves fundamentally differently. Using
\begin{equation}
\sum_{\gamma=1}^{3}\cos^6(\Theta-\phi_\gamma)=\frac{15}{16}+\frac{3}{32}\cos6\Theta,
\end{equation}
the energy of a uniform ferromagnetic state becomes
\begin{equation}
\frac{E_{\rm FM}(\Theta)}{N}
=
-\frac{3J_1}{2}
-\frac{9J_2}{8}
-J_3\left(\frac{15}{16}+\frac{3}{32}\cos6\Theta\right).
\label{eq:j1j2j3fmenergy}
\end{equation}
For $J_3>0$, the continuous degeneracy is therefore lifted already at zero temperature, and the energy is minimized by $\cos6\Theta=1$, giving $\Theta=n\pi/3$. In this case the sixfold anisotropy is explicit in the ground-state energy rather than generated by thermal fluctuations.

Figure~\ref{fig:higher_term_model_angle}(b) shows the corresponding angular distribution for $J_1=1$, $J_3=0.1$, and $J_2=0$. Although the same six lattice-aligned orientations are selected as in the $J_1$--$J_2$ model of panel (a), the peaks are considerably sharper and the probability between neighboring preferred directions is strongly suppressed. This difference reflects the distinct origin of the orientational locking. In the $J_1$--$J_2$ model, the sixfold anisotropy is entropic and arises only from the fluctuation free energy. By contrast, the $J_3$ term produces a finite energy difference between different values of $\Theta$, so the relative statistical weight of unfavorable orientations is increasingly suppressed as the temperature is lowered.

These results demonstrate that the preferred sixfold orientation is not fixed solely by the symmetry of the triangular lattice. Instead, both the selected directions and the mechanism responsible for their selection depend sensitively on the microscopic form of the bond-dependent interaction. The pure $J_1$ model selects $\theta=\pi/6+n\pi/3$ through thermal order-by-disorder, whereas adding a sufficiently strong $J_2$ term reverses the fluctuation-induced anisotropy and selects $\theta=n\pi/3$ without lifting the zero-temperature degeneracy. A positive $J_3$ term selects the same lattice-aligned orientations directly through the ground-state energy. 

The higher-order models therefore provide a useful bridge between fluctuation-induced and explicit energetic anisotropy, demonstrating two symmetry-allowed mechanisms that can produce lattice-aligned sixfold orientational order. Determining which, if either, is realized in the experimental Janus array requires a microscopic calculation of the interparticle interaction.

\section{Discussion and Outlook}

We have investigated the orientational ordering of Janus-particle arrays using anisotropic classical spin models motivated by the bond-dependent interactions of metallodielectric Janus particles. In the triangular-lattice $120^\circ$ model, thermal order-by-disorder lifts an accidental continuous ground-state degeneracy and selects six ordered states. The effective Kern--Frenkel model, despite its very different microscopic form, also retains a continuously degenerate ferromagnetic manifold and exhibits sixfold selection through thermal fluctuations. Both models show similar thermodynamic behavior, with a low-temperature sixfold ordered phase and a high-temperature disordered phase separated by an intermediate quasi-long-range-ordered regime consistent with two BKT transitions. The associated vortex--antivortex excitations provide a direct connection to the topological defects 
coincide with those observed experimentally~
\cite{park2025}.

Higher-order extensions of the $120^\circ$ interaction further reveal distinct mechanisms of orientational selection. A quadratic term can reverse the fluctuation-induced sixfold anisotropy while preserving the continuous ground-state degeneracy, whereas a cubic term lifts the degeneracy directly and selects the lattice-aligned orientations energetically. More broadly, these results show how different microscopic interactions can produce both fluctuation-induced and energetic sixfold order while retaining emergent XY-like behavior at intermediate scales.

A natural next step is to develop a more quantitative microscopic description of the effective interaction. The models studied here are deliberately simplified: the $120^\circ$ model isolates the leading compass-type bond anisotropy, the effective Kern--Frenkel model provides a geometrical representation of the orientation-dependent surface interaction, and the higher-order extensions illustrate how nonlinear bond-dependent couplings can alter both the mechanism and direction of orientational selection. For the metallodielectric particles used experimentally, however, the interaction ultimately originates from the spatially nonuniform polarization and induced-charge distributions generated by the AC electric field. Direct electromagnetic calculations of two-particle interactions, including their dependence on particle orientation, separation, field frequency, and electrolyte properties, would provide a route toward a quantitatively derived effective spin Hamiltonian. Such interactions need not reduce to a simple bilinear compass form and may naturally generate nonlinear or more complex bond-dependent terms of the type explored here. Janus arrays could therefore provide a platform not only for realizing known classical compass models, but also for exploring anisotropic spin interactions that are difficult to realize in conventional magnetic materials.

Nonequilibrium dynamics provide another important direction. The ability to image individual particle orientations in real time makes it possible to follow domain formation, vortex motion and annihilation, and relaxation directly following changes of the external field. A realistic dynamical description should account for rotational Brownian motion together with field-dependent interparticle interactions and possible electrohydrodynamic effects. Experimentally, varying the field amplitude changes the ratio of interaction energy to thermal energy, thereby providing direct control over the effective coupling strength. Rapid changes of this control parameter offer a natural setting for studying nonequilibrium ordering and defect dynamics. Preliminary experiments on the Janus arrays also suggest slow and possibly glassy relaxation following sudden changes in the applied field, motivating systematic studies of aging, defect kinetics, and nonequilibrium ordering.

In particular, the direct visualization and tracking of vortices and antivortices make Janus-particle arrays an attractive platform for studying the Kibble--Zurek mechanism (KZM), which relates topological-defect production to the rate at which a system is driven through a continuous phase transition \cite{kibble1976,zurek1985,delcampo2014}. Kibble--Zurek scaling has already been investigated experimentally in two-dimensional colloidal monolayers \cite{deutschlander2015}, demonstrating the utility of colloidal systems for resolving nonequilibrium domain and defect formation in real space. In the present system, the BKT character of the transitions makes vortex dynamics particularly important; studies of quenches in the two-dimensional XY model have shown that vortex annihilation and post-transition coarsening can modify the simplest Kibble--Zurek scaling picture~\cite{biroli2010,jelic2011}.


Finally, Janus-particle arrays may provide a setting to explore the Kibble--Zurek mechanism as a function of relaxation dynamics. In conventional KZM, faster passage through a continuous transition generally produces a higher density of topological defects~\cite{kibble1976,zurek1985,delcampo2014}. Combining real-space and real-time imaging in Janus-particle arrays with theoretical modeling and simulations offers an opportunity to investigate this interplay between relaxation dynamics and topological-defect formation. 

\acknowledgments
Z.~F. is supported by the Quantum Science and Technology-National Science and Technology Major Project (under Grant No.~2021ZD0301900) and the National Natural Science Foundation of China (NSFC) under Grant No.~12504265. M.~P. acknowledges support from the National Science Foundation under Grant No. DMR-2214590. M.~V. acknowledges support from the National Science Foundation under Grant No.~DMR-1944539. G.~W.~C. was partially supported by the U.S. Department of Energy, Office of Science, Basic Energy Sciences, through DE-SC0026087.

\appendix

\section{Effective interaction of Janus particles}
\label{app:effective-interaction}

Here, we provide a simple microscopic motivation for the effective interactions considered in the main text and show how higher-order compass terms can arise from the finite spatial separation of the polarization response within a Janus particle.

An AC electric field along the $z$ axis induces distinct complex polarization responses in the metallic and dielectric hemispheres of a Janus particle. To isolate the part of this response that is odd under interchange of the two hemispheres, we represent it by two opposite out-of-plane dipoles displaced along the in-plane particle orientation,
\begin{equation}
\mathbf R_{i,\pm}=\mathbf R_i\pm a\mathbf S_i,
\qquad
\mathbf p_{i,\pm}=\pm p_\omega\hat{\mathbf z},
\end{equation}
where $a$ is the effective displacement of the polarization centers and the complex amplitude $p_\omega$ contains the frequency-dependent response to the applied AC field. This antisymmetric component has vanishing net dipole moment, so its leading long-distance multipole is an electric quadrupole.

After averaging over an AC cycle, the interaction between particles $i$ and $j$ can be written in the form
\begin{equation}
U_{ij}=\mathcal A_\omega\sum_{s,t=\pm1}st\,
G\!\left(\left|\mathbf r_{ij}+a(t\mathbf S_j-s\mathbf S_i)\right|\right),
\label{eq:app-pairenergy}
\end{equation}
where $\mathbf r_{ij}=\mathbf R_j-\mathbf R_i$, $G(r)$ is the radial interaction kernel between two out-of-plane induced dipoles, and $\mathcal A_\omega>0$ contains the cycle-averaged dipolar amplitude. In linear response, $\mathcal A_\omega\propto |p_\omega|^2$ and therefore scales as the square of the applied electric-field amplitude.

For $a/r_{ij}\ll1$, Eq.~\eqref{eq:app-pairenergy} can be expanded in powers of the displacement. The zeroth- and first-order contributions cancel upon summing over $s,t=\pm1$, and the leading nonvanishing term is
\begin{equation}
U_{ij}^{(2)}=-4\mathcal A_\omega a^2
(\mathbf S_i\cdot\nabla)(\mathbf S_j\cdot\nabla)G(r_{ij}).
\label{eq:app-leading}
\end{equation}
For a radial kernel, the directional derivatives can be decomposed as
\begin{equation}
(\mathbf S_i\cdot\nabla)(\mathbf S_j\cdot\nabla)G
=
\frac{G'}{r}\,\mathbf S_i\cdot\mathbf S_j
+
\left(G''-\frac{G'}{r}\right)
(\mathbf S_i\cdot\hat{\mathbf r}_{ij})
(\mathbf S_j\cdot\hat{\mathbf r}_{ij}).
\label{eq:app-radial-derivative}
\end{equation}
For the illustrative unscreened dipolar kernel $G(r)=r^{-3}$, this gives
\begin{equation}
U_{ij}^{(2)}
=
\frac{12\mathcal A_\omega a^2}{r_{ij}^{5}}
\left[\mathbf S_i\cdot\mathbf S_j-5q_{ij}\right],
\qquad
q_{ij}=(\mathbf S_i\cdot\hat{\mathbf r}_{ij})(\mathbf S_j\cdot\hat{\mathbf r}_{ij}).
\label{eq:app-leading-interaction}
\end{equation}
Thus the leading interaction contains an antiferromagnetic isotropic exchange together with a stronger ferromagnetic compass interaction. On a fixed nearest-neighbor lattice, it is natural to write the corresponding smooth effective Hamiltonian as
\begin{equation}
\mathcal H_{\rm eff}
=
J_{\rm XY}\sum_{\langle ij\rangle}\mathbf S_i\cdot\mathbf S_j
-
\sum_{\langle ij\rangle}
\left(
J_1q_{ij}+J_2q_{ij}^{\,2}+J_3q_{ij}^{\,3}+\cdots
\right),
\label{eq:app-effective-H}
\end{equation}
with $J_{\rm XY}>0$ and $J_1>0$. At the leading order in Eq.~\eqref{eq:app-leading-interaction}, $J_1/J_{\rm XY}=5$. Although the isotropic contribution is antiferromagnetic, the stronger compass term makes the bond-aligned parallel configuration energetically favorable. The numerical ratio $1/5$ should not be regarded as a quantitative prediction for the experimental system, since screening, unequal hemisphere polarizabilities, and electrohydrodynamic effects will renormalize the effective couplings.

The finite separation of the polarization centers also generates higher-order angular dependence. This is particularly transparent for a locally parallel pair, $\mathbf S_i=\mathbf S_j=\mathbf S$. In this case Eq.~\eqref{eq:app-pairenergy} reduces to
\begin{equation}
U_{\parallel}
=
\mathcal A_\omega
\left[
2G(r)-G(|\mathbf r+2a\mathbf S|)-G(|\mathbf r-2a\mathbf S|)
\right].
\end{equation}
Writing $q=(\mathbf S\cdot\hat{\mathbf r})^2$, the expansion contains successive powers $q$, $q^2$, and $q^3$ at second, fourth, and sixth order in $a/r$, respectively. The orientational part can therefore be organized as
\begin{equation}
U_{\parallel}=U_0-J_1q-J_2q^2-J_3q^3+\cdots,
\end{equation}
where $U_0$ is independent of orientation and each coefficient also receives corrections from higher orders.

For $G(r)=r^{-3}$, matching the leading contribution to each power gives
\begin{equation}
\frac{J_2}{J_1}\simeq\frac{21}{4}\left(\frac{2a}{r}\right)^2,
\qquad
\frac{J_3}{J_1}\simeq\frac{1001}{40}\left(\frac{2a}{r}\right)^4.
\label{eq:app-higher-ratios}
\end{equation}
These estimates illustrate how nonlinear compass terms arise naturally from a smooth microscopic interaction when the internal polarization structure of the particle is resolved. They are not intended as quantitative predictions for the close-packed experimental array. Away from the locally parallel sector, the microscopic expansion can also generate additional symmetry-allowed mixed terms. Equation~\eqref{eq:app-effective-H} should therefore be viewed as a minimal smooth effective Hamiltonian retaining the isotropic exchange and the first few powers of the bond-dependent variable $q_{ij}$.

\section{Harmonic orientational selection}
\label{app:harmonic-selection}

We analyze the fluctuation-induced orientational selection of the $J_1$--$J_2$ model by first setting $J_{\rm XY}=J_3=0$ and expanding about a uniform ferromagnetic state, $\theta_i=\Theta+\delta\theta_i$. Let $\mathbf a_\gamma$ denote the three independent nearest-neighbor vectors of the triangular lattice, with bond angles $\phi_\gamma=0,\pi/3,2\pi/3$. To quadratic order,
\begin{equation}
\mathcal H^{(2)}
=
\frac{1}{2}\sum_{\mathbf r,\gamma}
B_\gamma(\Theta)
\left(
\delta\theta_{\mathbf r}
-
\delta\theta_{\mathbf r+\mathbf a_\gamma}
\right)^2,
\label{eq:app-harmonic-realspace}
\end{equation}
where
\begin{equation}
B_\gamma(\Theta)
=
\sin^2(\Theta-\phi_\gamma)
\left[
J_1+4J_2\cos^2(\Theta-\phi_\gamma)
\right].
\label{eq:app-Bgamma}
\end{equation}
Thus, although the classical energy of a uniform ferromagnet is independent of $\Theta$, the stiffness of its fluctuations depends on the global orientation.

After Fourier transformation,
\begin{equation}
\lambda_\Theta(\mathbf k)
=
2\sum_{\gamma=1}^{3}
B_\gamma(\Theta)
\left[
1-\cos(\mathbf k\cdot\mathbf a_\gamma)
\right],
\label{eq:app-lambda}
\end{equation}
and, excluding the uniform zero mode, the harmonic free-energy difference is
\begin{equation}
F_{\rm fl}(\Theta)-F_{\rm fl}(0)
=
\frac{T}{2}\sum_{\mathbf k\ne0}
\ln\frac{\lambda_\Theta(\mathbf k)}{\lambda_0(\mathbf k)}.
\label{eq:app-harmonic-freeenergy}
\end{equation}
The effect of $J_2$ is already evident by comparing the two interlaced sixfold sets: for $\Theta=0$, $(B_1,B_2,B_3)=\left(0,\frac34(J_1+J_2),\frac34(J_1+J_2)\right)$, whereas for $\Theta=\pi/6$, $(B_1,B_2,B_3)=\left(\frac14(J_1+3J_2),J_1,\frac14(J_1+3J_2)\right)$, up to a permutation of bond directions.

Numerical integration of Eq.~\eqref{eq:app-harmonic-freeenergy} over the Brillouin zone gives a reorientation point at
\begin{equation}
\frac{J_2}{J_1}\simeq0.0323.
\end{equation}
For smaller $J_2/J_1$, thermal order-by-disorder favors $\Theta=\pi/6+n\pi/3$, while for larger $J_2/J_1$ it favors $\Theta=n\pi/3$. The quadratic compass term therefore reverses the sign of the fluctuation-generated sixfold anisotropy without lifting the continuous degeneracy of the classical uniform ferromagnetic manifold. Strictly at $T=0$ the manifold remains continuous; the discrete selection refers to the equilibrium $T\to0^+$ Gibbs measure.

The isotropic exchange generated together with the leading compass term does not by itself select an absolute spin direction, since it contributes the same energy to every uniform ferromagnet. For the nearest-neighbor model with $J_3=0$ and $J_2\ge0$, the energy obeys
\begin{equation}
\mathcal H-E_{\rm FM}
\ge
\frac{J_1-4J_{\rm XY}}{8}
\sum_{\langle ij\rangle}
|\mathbf S_i-\mathbf S_j|^2,
\label{eq:app-fm-bound}
\end{equation}
with $E_{\rm FM}/N=3J_{\rm XY}-3J_1/2-9J_2/8$. Hence the microscopic estimate $J_{\rm XY}/J_1=1/5$ lies within the regime where the continuously degenerate ferromagnetic manifold is preserved. At harmonic order, including $J_{\rm XY}$ amounts to replacing $B_\gamma(\Theta)$ by $B_\gamma(\Theta)-J_{\rm XY}$. For $J_{\rm XY}/J_1=1/5$, the selected directions remain $\Theta=\pi/6+n\pi/3$ at $J_2=0$ and $\Theta=n\pi/3$ at $J_2/J_1=0.1$, while the reorientation point shifts to $J_2/J_1\simeq0.0527$. The value $0.0323$ quoted in the main text corresponds to the pure compass case $J_{\rm XY}=0$ used in the simulations.

\bibliography{ref}

\end{document}